\documentstyle[12pt,epsfig]{article}

\input epsf

\newcommand{\ppp}[1]{%
        \setbox0=\hbox{#1}%
        \kern-.02em\copy0\kern-\wd0
        \kern+.04em\copy0\kern-\wd0
        \kern-.02em\raise.0217em\box0}

\newcommand{\lsim}{
 \mathrel{\setbox0=\hbox{$<$}\raise0.6ex\copy0\kern-\wd0
 \lower0.65ex\hbox{$\sim$}}}

\newcommand{\gsim}{
 \mathrel{\setbox0=\hbox{$>$}\raise0.6ex\copy0\kern-\wd0
 \lower0.65ex\hbox{$\sim$}}}

\newcommand{\etal}{{\em et al.}}

\newcommand{\PRD}[3]{Phys.\ Rev.\ D {\bf {#1}}, {#2} ({#3})}
\newcommand{\PRC}[3]{Phys.\ Rev.\ C {\bf {#1}}, {#2} ({#3})}
\newcommand{\PRL}[3]{Phys.\ Rev.\ Lett.\ {\bf {#1}}, {#2} ({#3})}
\newcommand{\NPA}[3]{Nucl.\ Phys.\ A {\bf {#1}}, {#2} ({#3})}

\newcommand{\PLB}[3]{Phys.\ Lett.\ B {\bf {#1}}, {#2} ({#3})}

\newcommand{\JPG}[3]{J. Phys.\ G {\bf {#1}}, {#2} ({#3})}
\newcommand{\EJP}[3]{Eur. J. Phys.\ C {\bf {#1}}, {#2} ({#3})}

\begin{document}
%

\begin{titlepage}
\renewcommand{\thefootnote}{\fnsymbol{footnote}}
\makebox[2cm]{}\\[-1in]
\begin{flushright}
\begin{tabular}{l}
TUM/T39-98-33
\end{tabular}
\end{flushright}
\vskip0.4cm
\begin{center}
  {\Large\bf  Hard
     Exclusive Electroproduction of Pions\footnote{Work supported in
    part by BMBF}
}\\ 

\vspace{2cm}
L. Mankiewicz$^{a b}$, G. Piller$^a$ and A. Radyushkin$^{c d}$ \footnotemark\\

\vspace{1.5cm}

\begin{center}

{\em$^a$Physik Department, Technische Universit\"{a}t M\"{u}nchen, 
D-85747 Garching, Germany}

{\em $^b$N. Copernicus Astronomical Center, ul. Bartycka 18,
PL--00-716 Warsaw, Poland}

{\em $^c$Physics Department, Old Dominion University,
 Norfolk, VA 23529, USA}
 
{\em $^d$Jefferson Lab, Newport News,VA 23606, USA}

\footnotetext{Also Laboratory of Theoretical  Physics, 
JINR, Dubna, Russian Federation}

\end{center}

\vspace{1cm}


\vspace{3cm}

\centerline{\bf Abstract}
\begin{center}
\begin{minipage}{15cm}
We investigate the exclusive electroproduction of $\pi^+$ mesons 
from nucleons. 
To leading twist, leading order $\alpha_s$ accuracy the 
corresponding production amplitude can be decomposed into 
pseudovector and pseudoscalar parts. 
Both can be expressed in terms of  quark double distribution 
functions of the nucleon. 
While the pseudovector contribution is connected to ordinary 
polarized quark distributions, 
the pseudoscalar part can be related to one pion $t$-channel exchange.
We observe that the pseudovector part of the production amplitude 
is important at $x_{\rm Bj} < 0.1$.
On the other hand, for  $0.1< x_{\rm Bj} < 0.4$ 
contributions from one pion exchange dominate. 
\end{minipage}
\end{center}

\end{center}
\end{titlepage}
\setcounter{footnote}{0}

\newpage

\section{Introduction}

Hard exclusive electroproduction of mesons from nucleons has become 
a field of growing interest. 
The  recent progress in the QCD analysis of such processes 
is based on a factorization theorem \cite{CFS97} 
(see also \cite{Rad97,Hood98,MPW98a,Vanderh98a}) 
valid in the case of longitudinally polarized photons: 
at large photon virtualities, $Q^2 \gg \Lambda_{\rm{QCD}}^2$, 
the underlying photon-parton sub-processes are dominated by 
short distances and,
hence, 
 can be calculated perturbatively. 
On the other hand, all information on the long distance dynamics of 
quarks and gluons can be collected in the distribution 
amplitude of the produced meson  and in generalized 
(skewed \cite{Rad97,Ji98} or double \cite{Rad97}) 
parton distribution functions of the nucleon. 
As a consequence, experimental investigations
of meson production 
processes allow to probe details of the quark-gluon dynamics in 
nucleons beyond our current knowledge obtained
mainly from inclusive high-energy processes. 
Ongoing and planned measurements at DESY
(HERA \cite{HERA}, HERMES \cite{HERMES}),  
CERN (COMPASS \cite{COMPASS}), and Jefferson Lab \cite{TJNAF} 
are therefore of great interest.

In 
our 
previous work 
\cite{MPW98a,MPW98b}  we have derived 
the leading order QCD production amplitudes for neutral  
pseudoscalar mesons and vector mesons.  
The corresponding cross sections have been discussed within 
a specific model 
\cite{Rad98a} 
 for
double 
parton distributions.
In this paper  we consider hard electroproduction of charged pions. 
Interest in this process arises in particular from the fact that 
it receives significant contributions from two
production mechanisms: 
either (i) the  photon interacts with a quark  from the nucleon 
which, after hard gluon exchange, combines with a second quark 
of the target to the final pion, or (ii) the pion is produced 
from the meson cloud of the nucleon. The latter mechanism is often referred 
to as the pion pole contribution. 
In meson-cloud models of the nucleon  the second mechanism 
directly involves the electromagnetic form factor $F_{\pi}(Q^2)$ of 
the pion (see e.g. \cite{Speth} and references therein). 
Pion electroproduction has therefore often 
been considered as a generic process to determine  
$F_{\pi}(Q^2)$ (see e.g. \cite{bebek}). 
However, as discussed in ref.\cite{Carlson90}, 
there may be significant uncertainties 
due to  type (i) contributions. 
The latter have been modeled in ref.\cite{Carlson90}  
in terms of  hard gluon exchange diagrams with nonperturbative  
factors estimated  
by an overlap of light-cone wave functions  
with Chernyak-Zhitnitsky type distribution amplitudes \cite{CZ}. 

In this work   we address 
questions similar to those raised 
in \cite{Carlson90}.
The distinctive feature of our approach is a systematic use 
of perturbative QCD (PQCD) 
factorization  \cite{CFS97,Rad97}.
Within a consistent PQCD analysis  
all necessary information about the nucleon 
structure  is contained in   quark  
double 
distributions. 
The latter 
are constrained through the behavior of ordinary quark 
distribution functions, as well as through sum rules 
and symmetry properties (for a discussion see 
\cite{MPW98a,Rad98c}).
In particular we 
show that actually both production mechanisms, (i) and (ii), can 
be formulated in terms of 
quark double distribution functions  
of the nucleon, i.e.  the existence 
of the pion pole contribution fits into
the 
PQCD factorization framework. 

After modeling the involved double distribution functions   we 
calculate the differential $\pi^+$ production cross section 
in the region of small momentum transfers, $-t < 0.5$ GeV$^2$. 
We find that both mechanisms, 
(i) and (ii), contribute significantly. 
While one pion exchange dominates at intermediate 
values of Bjorken-$x$, perturbative quark exchange 
is relevant mainly at small $x_{\rm Bj}$.

The paper is organized as follows: in Sec.2 we present the 
amplitudes for charged pion electroproduction. Pseudovector and pseudoscalar 
parts of the relevant double distribution functions  
are associated in Sec.3 and 4 with the 
perturbative  and pion pole contribution, respectively.
Results are presented in Sec.5. Finally we 
summarize.

\section{Amplitudes}

The calculation of meson electroproduction amplitudes 
is based on   PQCD factorization  \cite{CFS97}. 
It applies to incident longitudinally polarized photons 
with large space-like momenta, $Q^2 = -q^2 \gg \Lambda_{\rm{QCD}}^2$,  
and moderate momentum transfer  to the nucleon target,  
$|t|\,\lsim  \Lambda_{\rm{QCD}}^2$.
Using the techniques  outlined in refs.\cite{CFS97,Rad97,MPW98a,MPW98b} 
gives the following 
result for the $\pi^+$ virtual photoproduction amplitude 
in leading twist, leading order $\alpha_s$ accuracy:
\begin{eqnarray}
\hspace*{-1cm}
{\cal A}_{\pi^{+}}
&=& i \frac{g^2 C_F}{4 N_c} \frac{f_{\pi}}{Q} 
           \frac{\bar{N}(P',S') \gamma_5 \hat{n} N(P,S)}{{\bar P} \cdot n}
           \int_0^1 d\tau \frac{\Phi_{\pi}(\tau)}{\tau \bar{\tau}}
          \int d\,[x,y] \, 
\nonumber \\
&&
\hspace*{-1.5cm}
\times  \left[ (e_d\, \Delta F^{du} - e_u\,\Delta {\bar F}^{du}) 
   \frac{{\bar \omega}}{x + 2 y + x \bar{\omega} - i \epsilon} - 
    (e_u \, \Delta F^{du} - e_d\, \Delta {\bar F}^{du})
    \frac{{\bar \omega}}{x + 2 y - x \bar{\omega} - i \epsilon} \right]  
    \nonumber \\
&&\hspace*{-1.5cm}
-i \frac{g^2 C_F}{2 N_c} \frac{f_{\pi}}{Q} 
           \frac{\bar{N}(P',S') \gamma_5  N(P,S)}{2 M}
           \int_0^1 d\tau \frac{\Phi_{\pi}(\tau)}{\tau \bar{\tau}}
          \int d\, [x,y] 
\nonumber \\
&&\hspace*{-1.5cm}
\times  \left[ (e_d\, \Delta K^{du} - e_u\,\Delta {\bar K}^{du}) 
   \frac{{1}}{x + 2 y + x \bar{\omega} - i \epsilon} - 
    (e_u \, \Delta K^{du} - e_d\, \Delta {\bar K}^{du})
    \frac{{1}}{x + 2 y - x \bar{\omega} - i \epsilon} \right].  
\nonumber \\
\label{eq:vecm_quark_diag}
\end{eqnarray}
Here the notation $\int d\, [x,y] \dots = \int_0^1 dx \int_0^{\bar x}
dy$ \dots,  with $\bar x = 1-x$ is used.  
$N(P,S)$ and $\bar N(P',S')$  are the  Dirac spinors   
of the initial and scattered nucleon, respectively, 
with the corresponding four-momenta  $P$, $P'$  
and spins $S$, $S'$. $M$ stands for the nucleon mass. 
The average nucleon momentum is denoted by $\bar P = (P + P')/2$, 
and the momentum transfer is  $r = P-P'$ 
with $t = r^2$. 
The produced  meson carries the four-momentum 
$q'$ and $\bar q = (q + q')/2$. 
Furthermore, we have introduced the variable 
$\bar \omega = 2 \bar q \cdot \bar P/ (- \bar q^2)$. 
Finally, $n$ is  a light-like vector with 
$n\cdot a = a^+ = a^0+ a^3$ \
for any vector $a$, and $\hat n = \gamma_{\mu} n^{\mu}$. 

The long distance dynamics of the 
pion in the final state 
is contained in the decay 
constant $f_{\pi} = 133$ MeV and the distribution amplitude 
$\Phi_{\pi}(\tau)$ \cite{ER78,BL79}:  
\begin{equation} \label{eq:meson_distr_VL}
\langle \pi^+ (q') |\, \bar{\psi}_u(x) \gamma_5\hat n \psi_d(y) 
\, | 0 \rangle
  =  - i f_{\pi} \,(q'\cdot n) \,\int_0^1  \, 
    \Phi_{\pi}(\tau) \,
    e^{i q' \cdot \, (\tau x + \bar{\tau} y)} \, d\tau \, ,  
\end{equation} 
where quark fields $\psi$ with proper flavor quantum numbers 
enter. 

The nucleon part of the amplitude (\ref{eq:vecm_quark_diag}) is determined 
by double distribution functions 
$\Delta F^{du}$ and $\Delta K^{du}$ 
which are nondiagonal in flavor.  
They are defined as matrix elements of a nonlocal quark 
operator sandwiched between proton and neutron states: 
\begin{eqnarray} \label{eq:dd_quark_ud}
&&\hspace{-1.5cm}
\left \langle n(P',S')\right| 
\bar\psi_d(0) \gamma_5\hat n \left[0,z\right] \psi_u(z) 
\left|p(P,S)\right \rangle_{z^2 = 0} 
= 
\nonumber \\
&&\bar N(P',S')\,\gamma_5 \hat n \,N(P,S) 
\int d\, [x,y] 
\left(
e^{-i x (P\cdot z) - i y(r\cdot z)} \Delta F^{du} 
+ 
e^{i x (P\cdot z) - i \bar y(r\cdot z)} \Delta \bar F^{du}
\right) 
\nonumber \\
&-& 
\bar N(P',S')\,\gamma_5 \,N(P,S) \,\frac{r \cdot n}{2 M}
\int d\, [x,y] 
\left(
e^{-i x (P\cdot z) - i y(r\cdot z)} \Delta K^{du} 
+  
e^{i x (P\cdot z) - i \bar y(r\cdot z)} \Delta \bar K^{du}
\right). 
\nonumber \\
\end{eqnarray}
Here, as in (\ref{eq:vecm_quark_diag}),  the dependence of 
the double distribution functions on $x, y$ and $t$ has been 
suppressed. 
The up- and down-quark fields in (\ref{eq:dd_quark_ud})   
are separated by a light-like distance $z\sim n$. 
Gauge invariance is guaranteed by the 
path-ordered exponential  
$$[0,z] = \,
{\cal P} \exp [ -i g z_{\mu} 
\-\int_0^1  A^{\mu}(z \lambda) \, d \lambda]$$  
which reduces to 
$1$ 
in  axial gauge $n\cdot A=0$ 
($g$ stands for the strong coupling constant and 
$A^{\mu}$ denotes the gluon field).

\section{Pseudovector contribution}

The production amplitude ${\cal A}_{\pi^+}$ contains 
a pseudovector and pseudoscalar contribution
proportional to  $\bar N(P',S')\,\gamma_5 \hat n \,N(P,S)$ 
and $\bar N(P',S')\,\gamma_5 \,N(P,S)$, respectively. 
We rewrite the pseudovector part according to isospin 
relations which connect  flavor non-diagonal double distribution 
functions $\Delta F^{du}$ with flavor diagonal ones. 
In particular we use \cite{MPW98b}:  
\begin{equation}
\left \langle n \right| {\hat O^{du}}(z) \left|p\right \rangle =   
\left \langle p\right| {\hat O^{uu}}(z) \left|p\right \rangle - 
\left \langle p\right| {\hat O^{dd}}(z) \left|p\right \rangle,
\end{equation}
with 
\begin{equation}
\left.{\hat O}^{q\,q^\prime}(z) =  
\bar\psi_q(0) \gamma_5\hat n \left[0,z\right]
\psi_{q^\prime}(z)\right|_{z^2 = 0}. 
\end{equation}

In terms of flavor diagonal polarized double distribution 
functions the pseudovector part of the amplitude 
(\ref{eq:vecm_quark_diag}) reads: 
\begin{eqnarray}
{\cal A}_{\pi^{+}}
&=& - i \frac{g^2 C_F}{8 N_c} \frac{f_{\pi}}{Q} 
           \frac{\bar{N}(P',S') \gamma_5 \hat{n} N(P,S)}{{\bar P} \cdot n}
           \int_0^1 d\tau \frac{\Phi_{\pi}(\tau)}{\tau \bar{\tau}}
           \int d\, [x,y] \Bigg{\{}
\nonumber \\
&& 
(e_u\! - \!e_d) 
   \left[(\Delta F^u \!+\! \Delta {\bar F}^u) -  (
          \Delta F^d\! +\! \Delta {\bar F}^d)\right] 
 \left(  \frac{{\bar \omega}}
{x\! + \!2 y\! +\! x \bar{\omega}\! - \!i \epsilon}\! +\!
\frac{{\bar \omega}}
{x \!+\! 2 y\! - \!x \bar{\omega}\! - \!i \epsilon} \right)
\nonumber \\
&-& 
(e_u \!+ \!e_d) 
   \left[(\Delta F^u \!- \!\Delta {\bar F}^u) -  
         (\Delta F^d \!-\! \Delta {\bar F}^d)\right] 
 \left(\frac{{\bar \omega}}
{x\! + \!2 y\! +\! x \bar{\omega} \!- \!i \epsilon} -
         \frac{{\bar \omega}}
{x \!+\! 2 y\! - \!x \bar{\omega} \!- \!i \epsilon} \right) 
\Bigg{\}}.
\nonumber \\
\label{eq:piplus_quark_diag}
\end{eqnarray}
Here $e_u$ and $e_d$ denote the electromagnetic charges of 
the up- and down-quarks.
Note that ${\cal A}_{\pi^{+}}$ 
has a structure similar to the $\rho^+$ production 
amplitude derived in \cite{MPW98b}: 
up to prefactors the  $\rho^+$ amplitude  can be  obtained 
from (\ref{eq:piplus_quark_diag}) by replacing polarized quark 
double distribution functions  by unpolarized ones,
with proper account of 
their opposite charge conjugation properties.

In order to obtain a numerical estimate  for the 
pseudovector contribution (\ref{eq:piplus_quark_diag}) 
to $\pi^+$ production models for the involved 
double distributions $\Delta F$ have to be constructed. 
We are guided here by the appropriate forward limit of 
double distributions, their symmetry properties, 
and sum rules which relate them to nucleon form factors.
Following refs. \cite{Rad98b} and \cite{Rad98c} we use: 
\begin{eqnarray} \label{eq:f_models}
A: \quad  \Delta F(x,y;t) & = & h(x,y) \, \Delta q(x) \, f(t), 
\nonumber \\
\nonumber \\
B: \quad \Delta F(x,y;t) & = & 
h(x,y) \, \Delta q(x) \, 
\exp\left[\frac{t}{\Lambda^2} \, \frac{y(1-x-y)}{x(1-x)}\right]. 
\end{eqnarray}
Here $h(x,y) = 6 \, y \,(1-x-y)/(1-x)^3$, 
and $\Delta q(x)$ denotes  the corresponding ordinary polarized 
quark distribution. 
  In numerical
  calculations presented in this paper we have used Gehrman-Stirling LO set-A
  parametrizations of polarized quark distributions \cite{Gehrmann96}.  
Furthermore,  
we have used the one loop expression for the running coupling constant  
with $\Lambda_{\rm QCD} = 200$ MeV. 
Note  that
in addition to 
the above mentioned constraints, 
the models (\ref{eq:f_models}) 
are consistent with  asymptotic solutions of evolution equations 
for double distributions  
and satisfy the symmetries of the latter \cite{MPW98a}. 
The $t$-dependence
of double distributions in  model $A$ is governed by the form factor
$f(t)$ which, like in \cite{MPW98a}, is taken to be equal to
the nucleon pseudovector form factor \cite{Weise88}:
\begin{equation}\label{eq:ffac_model_1}
f(t) = \frac{1}{(1-t/\Lambda^2)^2}.
\end{equation}
The exponential dependence of $\Delta F$ on 
the momentum transfer $t$ in model $B$ can be
obtained from a simple field theoretical  
investigation of double distribution
functions outlined in \cite{Rad98c,Rad98b}. 
For both models the scale $\Lambda$ has been  fixed at $1$ GeV. 
With this choice   model $B$ 
reproduces the nucleon axial form factor up to momentum transfers 
$t \approx - 0.5$ GeV$^2$. 
Another important feature of model $B$ is, that the small-$x$ 
behavior of $\Delta F$ becomes less and less singular as $|t|$ increases. 
Such a behavior is expected since, like in the case of nucleon
form factors, large momentum transfers filter out the minimal 
contribution to the Fock space wave function of the target.

\section{Pseudoscalar contribution}

As explained in ref.\cite{Rad98b}  the double 
distributions $\Delta F(x,y;t)$ can be related in the forward
limit, $t = 0$, to ordinary polarized quark densities $\Delta q(x)$. 
Therefore, it  is possible to construct plausible models for 
$\Delta F$,   
although little is known about their shape from first principles. 
On the other hand, for the double distributions $\Delta K$ 
no experimental information on  corresponding forward densities is available 
which makes modeling in this case more difficult. 
Nevertheless, some intuition about their magnitude has been obtained from 
model calculations \cite{Models}.
Still, 
contributions of the so-called $K$-terms 
to hard exclusive meson production have  been neglected
so far.  
In the present case of $\pi^+$ 
production the situation is  different. As  
we will show,  the pseudoscalar piece of the amplitude ${\cal A}_{\pi^+}$ 
can be associated with the virtual photoproduction of pions from 
the nucleon  pion cloud.  
Due to the small pion mass  this mechanism is expected to be
important, especially at small  momentum transfers $t$.

We are interested in evaluating the contribution to the 
nucleon matrix element
(\ref{eq:dd_quark_ud}) arising from a  situation when a quark-antiquark
pair with  low invariant mass propagates  in the 
$t$-channel, such that it becomes close to a  $\pi^+$ bound state.
To construct the corresponding double distribution function we 
introduce an effective Lagrangian describing a pseudovector pion-nucleon
interaction \cite{Weise88}:  
\begin{equation}
{\cal L}_{\pi NN} = 
\frac{g_{\pi NN}}{2 M} 
\left( \phi(x) \gamma_{\mu} \gamma_5 \vec\tau \phi(x)\right)
\cdot \left(\partial^{\mu} \vec \varphi(x)\right).
\end{equation}
The nucleon and pion fields are denoted by 
$\phi$ and $\vec \varphi$, respectively,  
$\vec \tau$ are  the common isospin matrices, 
and $g_{\pi N N}$ is the pion-nucleon coupling constant.  
Evaluating the matrix element (\ref{eq:dd_quark_ud}) to  
first order in $g_{\pi NN}$ leads to:
\begin{eqnarray} 
&&
\hspace*{-1cm}
\left \langle n(P',S')\right| 
\bar\psi_d(0) \gamma_5\hat n \left[0,z \right] \psi_u(z) 
\left|p(P,S)\right \rangle_{z^2 = 0} 
= \sqrt{2} g_{\pi NN} \,
\bar N(P',S')\,\gamma_5 \,N(P,S)
\nonumber \\
&&
\times \int d^4 x  \, e^{- i r \cdot x}
\left \langle 0 \right| 
T[
\bar\psi_d(0) \gamma_5\hat n \left[0,z \right] \psi_u(z) 
\, \phi^+(x) ]
\left| 0  \right \rangle_{z^2 = 0} \, ,
\label{eq:repl1}
\end{eqnarray}
where $\varphi^+ = \frac{1}{\sqrt{2}}(\varphi_1+i\, \varphi_2)$ is the pion
field associated with the  $\pi^+$ meson. The leading-twist contribution
to the matrix element (\ref{eq:repl1}) arises from the 
region of phase space where the quark fields are localized 
near the light-cone, $z^+ \sim 0$. 
Inserting a full set of intermediate pion states between the pion
field and the non-local quark operator gives:
\begin{eqnarray} \label{eq:ps_contr}
&&
\hspace*{-1cm}
\left \langle n(P',S')\right| 
\bar\psi_d(0) \gamma_5\hat n \left[0,z\right] \psi_u(z) 
\left|p(P,S)\right \rangle_{z^2 = 0} 
\nonumber \\ 
&&
=
N(P',S') \gamma_5 N(P,S)\, 
\frac{- i \sqrt{2} g_{\pi NN}}{m_\pi^2 - t}\,
\langle 0 | 
\bar\psi_d(0) \gamma_5\hat n \left[0,z\right] \psi_u(z)
|\pi^+(r)\rangle_{z^2 = 0}.
\end{eqnarray}
Using the pion distribution amplitude (\ref{eq:meson_distr_VL}) and 
comparing with the definition of the 
pseudoscalar double distribution functions in eq.(\ref{eq:dd_quark_ud}) 
yields: 
\begin{eqnarray}\label{eq:mel_ps_final} 
\left(\Delta K^{du}  + \Delta \bar K^{du}\right)(x,y,t\approx 0) 
&=& 
-\frac{2 \sqrt{2} f_{\pi} M g_{\pi NN}}{m_\pi^2 - t} 
\,\delta (x) \,\Phi_{\pi}(y). 
\end{eqnarray}
As this contribution can be expressed entirely in terms of $\Delta K$
there is no danger of double counting if we  add this term to the
contribution arising from a model for $\Delta F$, as discussed in the
previous section. 

One can express the $t$-channel one pion exchange
contribution also in the form:
\begin{equation}\label{eq:dir_pion_explicit}
{\cal A}_{\pi^{+}} =
-\sqrt{2} g_{\pi NN}
\frac{N(P',S') \gamma_5 N(P,S)}{m_\pi^2 - t}\,
\,\varepsilon_L \cdot (q^\prime + r) \,F_{\pi^+}(Q^2), 
\end{equation}
where
\begin{equation}\label{eq:e_L_def}
\varepsilon_L = \frac{i}{Q}\, (q^\prime + r)
\end{equation}  
stands for  the polarization vector of the longitudinally polarized photon, 
and 
\begin{equation}\label{eq:F_pi_Q2}
F_{\pi^+}(Q^2) = (e_u-e_d)\, \frac{g^2 C_F}{2 N_c} \frac{f_\pi^2}{Q^2} 
\left(\int_0^1 du\, \frac{\Phi(u)}{u}\right)^2 
\end{equation}
is the leading QCD contribution 
to the $\pi^+$ electromagnetic form factor at
large $Q^2$ \cite{ER78,BL79}.

To twist-2 accuracy, i.e. neglecting terms of  order $t/Q^2$ 
and $m_\pi^2/Q^2$, the amplitude 
(\ref{eq:dir_pion_explicit}) is explicitly gauge invariant. 
This is guaranteed by the factor $\varepsilon_L \cdot (q^\prime + r)$ 
which arises from  the hard photon-quark interaction.
In the one pion exchange approximation  the
difference $\Delta K^{du} - \Delta \bar K^{du}$ decouples from the considered
production process. 
Finally, note that in the case of $\pi^0$ production 
a $t$-channel exchange contribution similar to 
(\ref{eq:mel_ps_final}) appears in the production
amplitude with a zero coefficient from the difference 
of two equal quark charges.

To this end some comments regarding  one pion $t$-channel exchange, 
as used above, are in order. 
In general one expects that at large center-of-mass energies 
any $t$-channel exchange of an hadronic state should be 
replaced by the corresponding Regge trajectory. 
Therefore, at small values of Bjorken-$x$ additional contributions 
to the pseudoscalar part of the production amplitude 
${\cal A}_{\pi^+}$ could occur. 
Once the pion pole contribution is replaced by  reggeized pion (and rho)
exchange \cite{Guidal97} the relation to the pion 
electromagnetic form factor seems to be lost 
(see, however, ref.\cite{Vanderh98b} for an attempt to interpret pion 
electroproduction data in the framework of a  Regge exchange  model). 
We restrict ourselves in this work to the 
simplest model for the pseudoscalar part of the production 
amplitude in the hope that it contains the dominant 
contribution at small momentum transfers, i.e. close to the 
pion pole.

\section{Results}

In Figs.1 and 2 we present results for the  
cross section  of  exclusive $\pi^+$ production 
in the scattering of longitudinally polarized 
photons from nucleons  at $Q^2 = 10$ GeV$^2$. 
Up to logarithmic corrections the  cross 
section is proportional to $1/Q^6$ as one can infer from 
eq.(\ref{eq:vecm_quark_diag}).

In Fig.1 we show the differential production cross section 
for the minimal kinematically 
allowed value of $t = t_{\rm min} = -x_{\rm Bj}^2 M^2/(1-x_{\rm Bj})$.
We find that the pseudovector contribution (\ref{eq:piplus_quark_diag}) 
dominates at small values of the Bjorken variable $x_{\rm Bj}$. 
Its $x_{\rm Bj}$-dependence reflects the behavior of 
the relevant 
forward parton distributions (\ref{eq:f_models}).
At $x\sim 0.3$ the cross section is governed by the pseudoscalar  
pion cloud contribution 
(\ref{eq:ps_contr}).

While models $A$ and $B$ give qualitatively similar cross sections
at small momentum transfers,  they start to differ for increasing  $|t|$. 
This is illustrated in Fig.2 
where we present results for the differential production 
cross section taken at $t = - 0.4$ GeV$^2$. 
The maximum value of Bjorken-$x$ is determined by the condition 
$|t| \geq |t_{\rm min}|$. 
For model $A$ the pseudovector contribution decreases slowly 
with  decreasing $x_{\rm Bj}$ and dominates the cross 
section at $x_{\rm Bj} \lsim  0.1$. 
At $x_{\rm Bj} \sim x_{\rm max}$ the total 
pion production cross section is about half as large as   
obtained from the pion cloud contribution alone due to 
strong interference effects. 
On the other hand, in model $B$ the $x_{Bj}$-dependence of 
the pseudovector contribution is similar to the one 
from the pion cloud. 
Also, contrary to the previous case, at 
$x_{\rm Bj} \sim x_{\rm max}$ the production cross 
section is about  20\% larger than 
obtained from the pion cloud alone.

Finally we compare  in Fig.3  to the leading twist 
production cross section for $\pi^0$ mesons  
which has been investigated  in \cite{MPW98a}. 
At $t=t_{\rm min}$ the pseudovector contribution to $\pi^+$ 
production is about a factor $5$--$10$ larger as compared to 
the $\pi^0$ case. 
At $x_{\rm Bj}\sim 0.3$ large contributions from the pion cloud, 
which are absent in $\pi^0$ production, add further 
to the cross section ratio.

It should be emphasized that the dominance of 
the hard gluon exchange processes considered in this  
paper can only be ensured for sufficiently large values of $Q^2$. 
Just like in the case of the pion form factor, 
competing contributions  are provided by 
soft mechanisms  which, in this case, correspond   
to the nonperturbative overlap of three hadronic wave functions. 
The interplay of soft and hard interaction mechanisms 
has been studied  recently  for the pion form factor in a  
QCD sum rule related model \cite{SJR98}. 
Guided by the results of this investigation we expect 
the  dominance of 
hard gluon exchange for values of 
$Q^2$ above 10 GeV$^2$.

\section{Summary}

We have discussed the hard exclusive electroproduction of $\pi^+$ mesons 
from nucleons. 
For longitudinally polarized photons one can 
express the corresponding photoproduction amplitude 
to leading twist, leading order $\alpha_s$ accuracy 
in terms of quark double distribution functions of the nucleon. 
We find that the amplitude receives contributions 
from pseudovector and pseudoscalar pieces. 
The pseudovector contribution is determined by quark 
double distributions with well known 
properties in the forward limit, as given by  
ordinary polarized quark distribution functions.
However, for the pseudoscalar part such information 
is not available. 
In this work it has been modeled in terms of an interaction of 
the virtual photon with the nucleon pion cloud.
As a consequence, the related double distribution is determined 
by the pion distribution amplitude. 
The corresponding contribution to the pion production
amplitude turns out to be proportional to the pion form factor.

We have found that both,  pseudovector and pseudoscalar  
terms
should be included if one considers the whole $0 <x_{\rm Bj} <1$ range. 
In particular, 
we have observed 
that the pseudovector part dominates at small 
$x_{\rm Bj} < 0.1$. 
On the other hand, the pion cloud contribution controls the 
$\pi^+$ production 
at small $t \sim t_{\rm min}$ and  $0.1 < x_{\rm Bj} < 0.4$. 
At larger values of $|t| \sim 0.4$ GeV$^2$ the relative weight 
of pseudoscalar and pseudovector contributions 
has been found to be sensitive to details of the
relevant 
double 
distribution functions.
Further investigations of double distribution functions, 
as well as experimental data on exclusive meson production 
are certainly needed.

\bigskip
 
\noindent
{\bf Acknowledgments:} 

\noindent
G.P. would like to thank the SLAC theory group for their hospitality. 
Furthermore, we would like to acknowledge discussions with 
V. Burkert, S.J. Brodsky, L. Frankfurt, M.V. Polyakov and P. Stoler.
The  work of A.R.  was supported by the 
US Department of Energy under contract 
DE-AC05-84ER40150. 
\\
\noindent
After this paper has been completed, we learned about an
independent calculation of $\pi^+$ electroproduction 
which uses skewed parton distributions computed in a 
chiral quark soliton model 
\cite{Bochum1}. 
The corresponding results turn out to be quite similar to ours.
We thank the authors of \cite{Bochum1} 
for discussions and an exchange of results.

\newpage


\begin{figure}
\begin{center} 
\hspace*{0.4cm} 
\epsfig{file=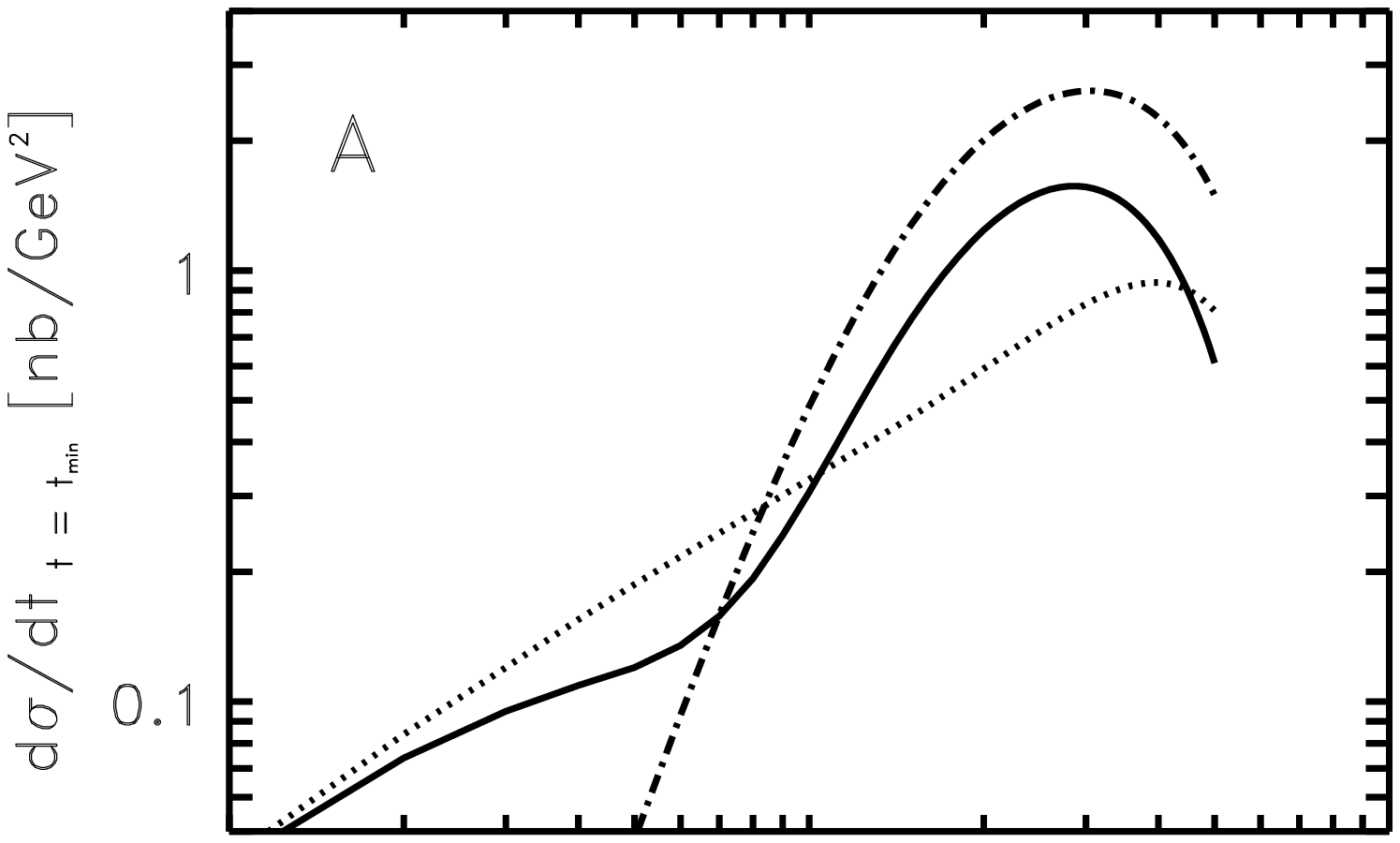,height=70mm,width=106mm}
\end{center}
\end{figure}
 
\begin{figure}
\vspace*{-3.cm} 
\begin{center} 
\hspace*{1cm} 
\epsfig{file=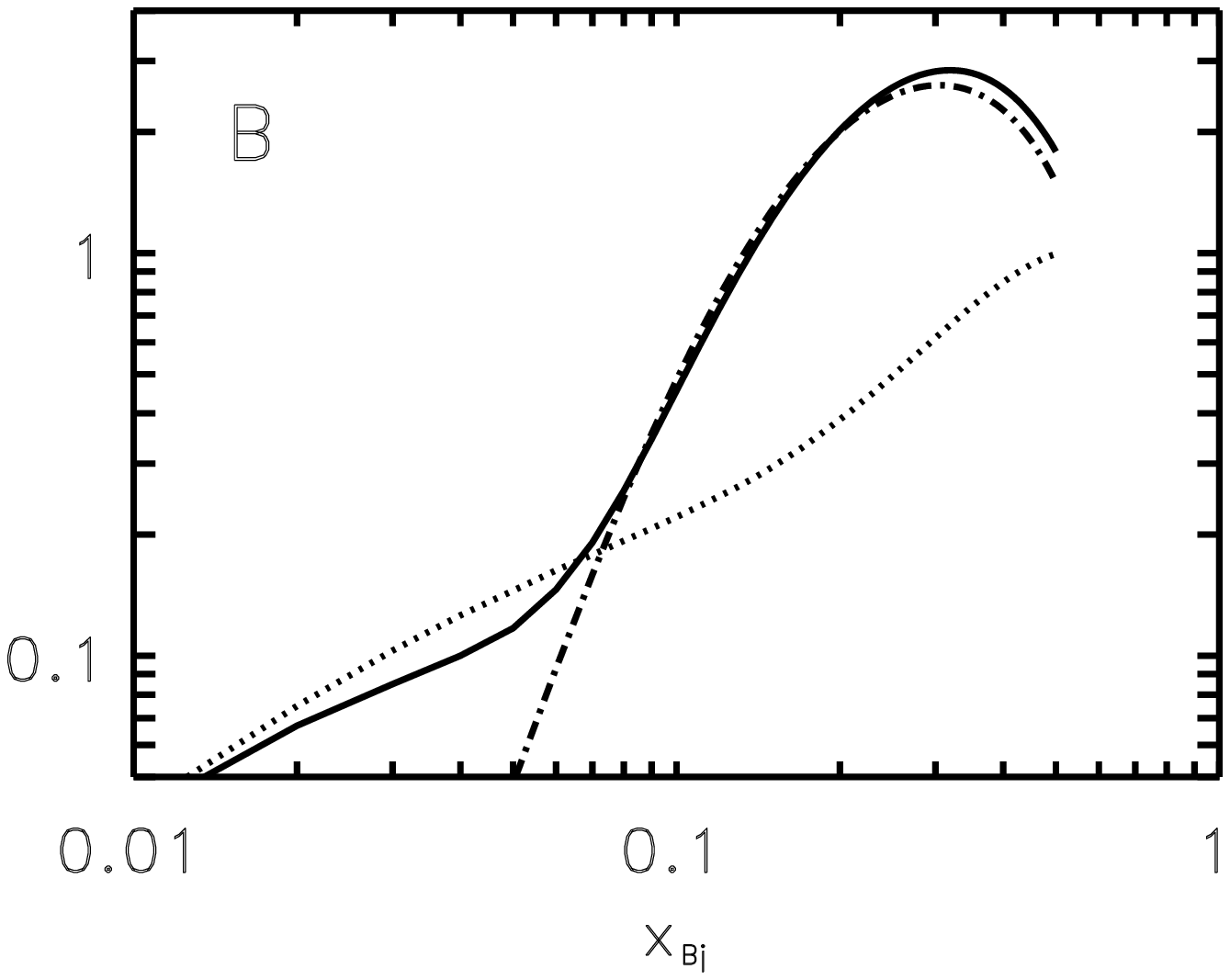,height=90mm,width=100mm}
\end{center}
\caption[...]{ 
Differential production cross section for exclusive $\pi^+$ 
production through the scattering of longitudinally polarized 
photons from protons at $t=t_{\rm min}$ and $Q^2 = 10$ GeV$^2$. 
The dotted and dot-dashed curves show  the pseudovector and 
pseudoscalar contributions, respectively. 
The upper figure corresponds to model $A$, the lower one to model $B$. 
}
\end{figure}

\pagebreak

\begin{figure}
\begin{center} 
\hspace*{0.6cm} 
\epsfig{file=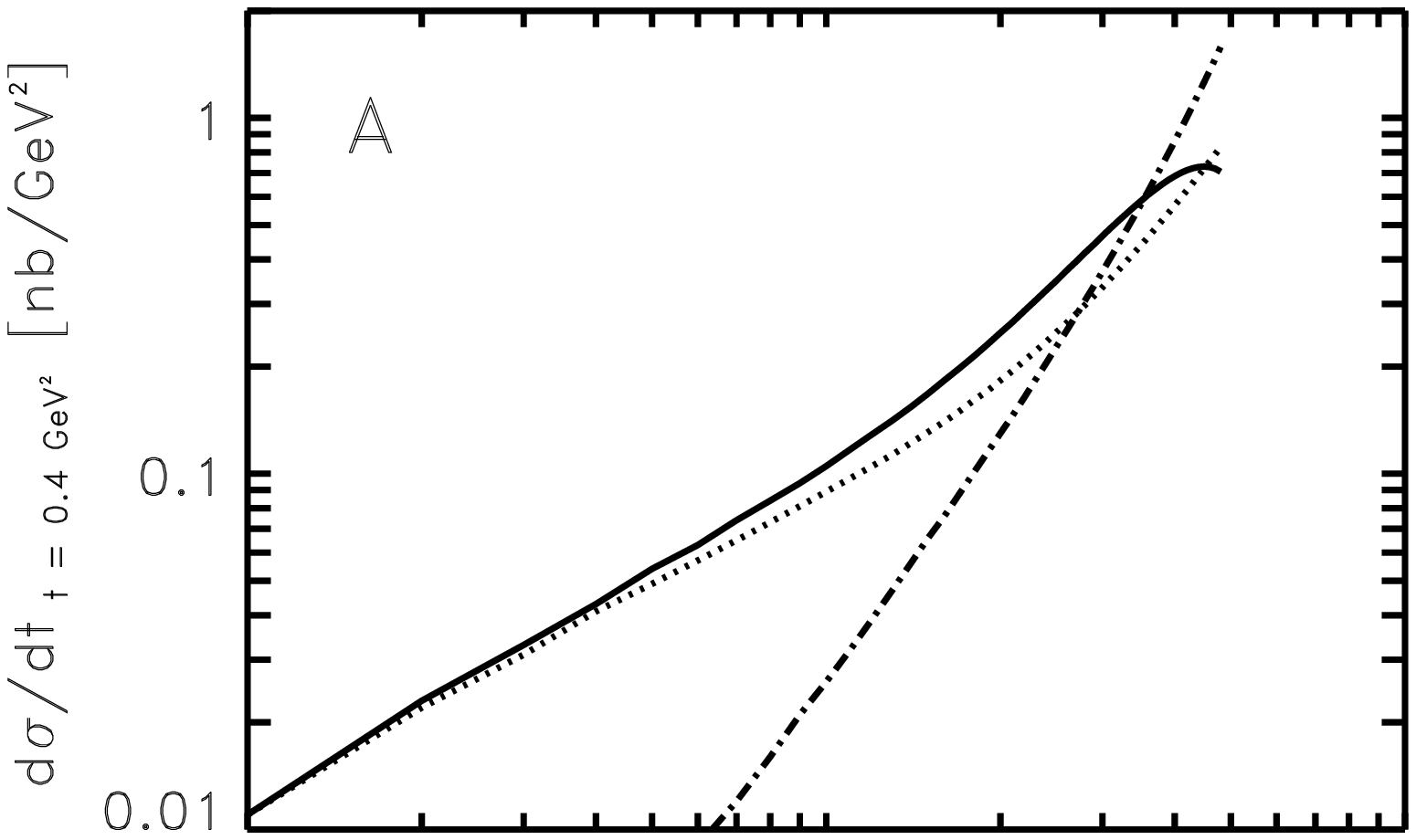,height=70mm,width=105mm}
\end{center}
\end{figure}
 
\begin{figure}
\vspace*{-3.cm} 
\begin{center} 
\hspace*{1cm} 
\epsfig{file=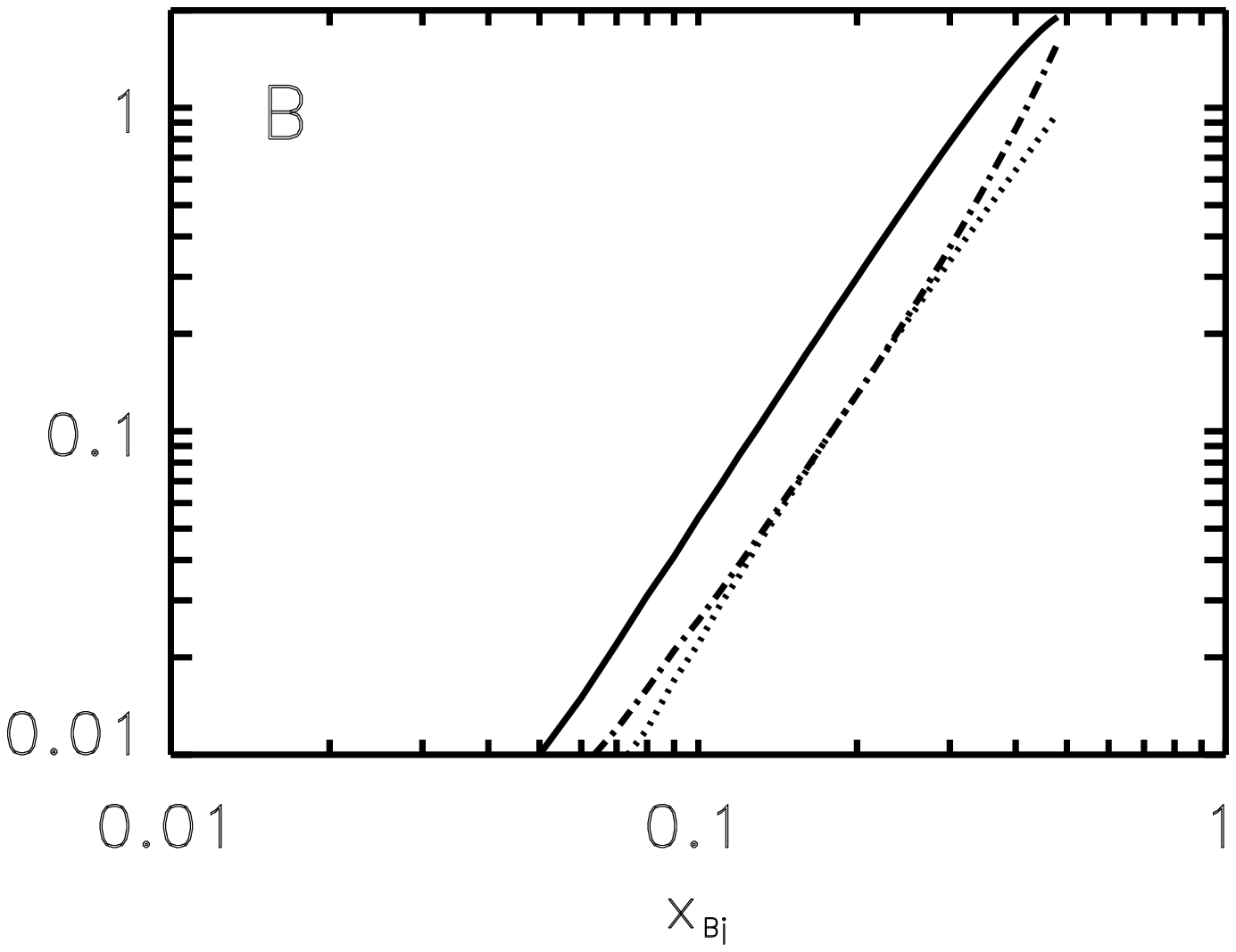,height=90mm,width=100mm}
\end{center}
\caption[...]{ 
Differential production cross section for exclusive $\pi^+$ 
production through the scattering of longitudinally polarized 
photons from protons at $t=-0.4$ GeV$^2$ and $Q^2 = 10$ GeV$^2$ . 
The dotted and dot-dashed curves show the pseudovector and 
pseudoscalar contributions, respectively. 
The upper figure corresponds to model $A$, the lower one to model $B$. 
}
\end{figure}

\pagebreak

\begin{figure}
\label{fig:ST}
\vspace*{2.cm} 
\begin{center} 
\hspace*{1cm} 
\epsfig{file=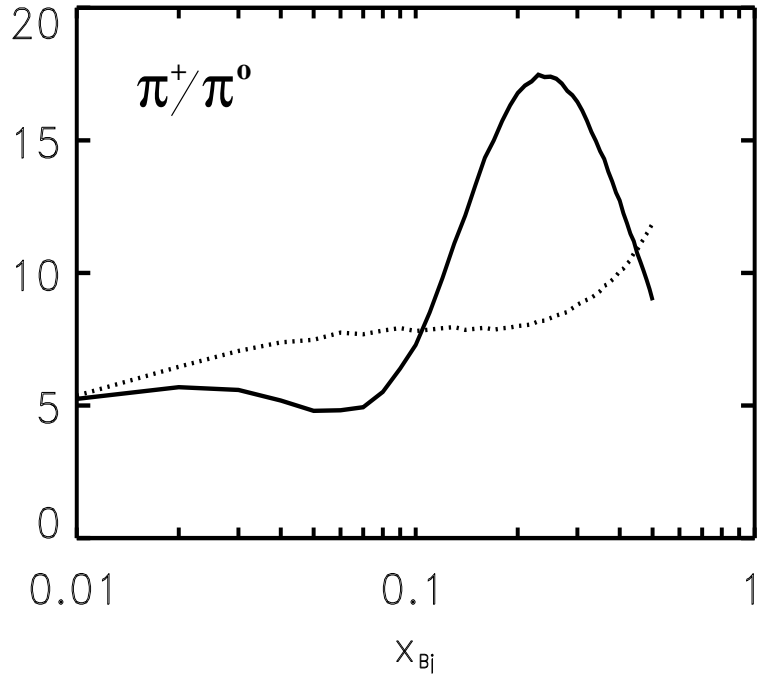,height=90mm,width=100mm}
\end{center}
\caption[...]{ 
Ratio of the differential production cross sections for exclusive $\pi^+$ 
and $\pi^0$ production through the scattering of longitudinally polarized 
photons from protons at $t=t_{\rm min}$  and $Q^2 = 10$ GeV$^2$ 
for model $A$. 
The dotted curve  corresponds to the pseudovector contribution alone.
}
\end{figure}

\end{document}